\documentclass[pra,twocolumn]{revtex4}

\usepackage{graphicx}
\usepackage{amssymb}

\begin{document}

\title[Asymmetric Foucault pendulum analogies to the Lipkin-Meshkov-Glick model]{Asymmetric Foucault pendulum dynamics with analogies to the Lipkin-Meshkov-Glick quantum phase transitions and other quantum phenomena}
 \author{Tom\'{a}\v{s} Opatrn\'{y}$^1$  and Pavel \v{S}t\v{e}p\'{a}nek$^2$}
 \address{$^1$Optics Department, Faculty of Science, Palack\'{y} University, 17. Listopadu 12,
 77146 Olomouc, Czech Republic\\
 $^2$ Mendel Gymnasium, Komensk\'{e}ho 5,
746 01 Opava, Czech Republic}
%\ead{opatrny@optics.upol.cz}

\begin{abstract}
Stokes parameter formalism is applied to show the analogies between the motion of an asymmetric Foucault pendulum and several phenomena known from optics and atomic physics. Nonlinearity-induced precession of elliptical orbits of the pendulum is shown to correspond to twisting transformations used for spin squeezing of atomic systems. Transitions between regimes of predominant nonlinearity and regimes where the Coriolis force or the asymmetry of the pendulum are dominant correspond to quantum phase transitions in the Lipkin-Meshkov-Glick model. A Foucault pendulum with highly anisotropic damping can emulate an optical Zeno effect where a sequence of polarizing filters inhibits polarization rotation of light in an optically active medium.
\end{abstract}

\date{\today }

 \maketitle

\section{Introduction}
%%%%%%%%%%%%%%%%%%%%%%%%%%%%%%%%%%%%%%%%%%%%%%%%%%%%%%%%%%%%%%%%%%%%%%%%%%%%%%
To check whether Earth is still rotating, one can suspend a weight on a rope, start it swinging and observe the precession of its trajectory. The plane of swinging should precess with rate $\Omega_E \sin \lambda$, where $\Omega_E$ is the Earth's rotational angular velocity and $\lambda$ is the geographic latitude. However, replicating the original Foucault's pendulum experiment of 1851 is not as easy as it sounds. Anybody who tried it for the first time has probably noticed a strange behavior: 
%a trajectory which has initially straight-line horizontal projection gradually 
the horizontal projection of the trajectory changes from being a straight line to be approximately elliptical.
%with originally straight horizontal projection gradually becomes elliptical-like. 
Elliptical trajectories exhibit precession that is far from what one would naively expect considering just Earth's rotation. These problems have been studied since the days of the earliest experiments with Foucault's pendulums. In 1851 Airy showed that elliptical trajectories precess due to the nonlinear character of the spherical pendulum, the precession rate being proportional to the area of the ellipse \cite{Airy} (for more recent derivations see \cite{Olsson,Olsson-1981,Maya}). Kamerlingh-Onnes in his dissertation of 1879 \cite{Onnes} studied a finely tuned Foucault's pendulum and showed that even a small asymmetry in the pendulum's construction leads to a transition from a straight trajectory to elliptical (for a discussion of Kamerlingh-Onnes' contribution in English see \cite{Schulz}). 
The primary observation was that
 the motion consists of perpendicular oscillations of different frequencies with the accumulated phase difference changing the trajectory shape.

A complete analytical description of the anisotropic spherical pendulum in a rotating system is rather involved and various aspects have been covered in extensive literature (see, e.g., \cite{Maya, Johansen,Holm, Moeckel-2017}). Here we present a simple intuitive picture that describes most of the important phenomena, based on the Stokes parameter formalism and the Poincar\'{e} sphere visualization. 
These tools are typically used to deal with the
dynamics of polarized light or spin systems but seldom have been applied to mechanical motion (for a recent exception, see \cite{Verreault-2017}).
Surprisingly, the pendulum motion then can be mapped to various phenomena that appear in completely different areas of physics. 
In particular, the Airy precession corresponds to the ``one-axis-twisting'' known from the physics of spin squeezing \cite{Kitagawa,Esteve2008,Gross2010,Riedel,Leroux-2010,Opatrny2015}, whereas precession of the orbit due to the Coriolis force corresponds to the energy splitting, and the asymmetry-driven dynamics noted by Kamerlingh-Onnes  to Rabi oscillations. Combining these effects, one finds behavior analogous to  transitions between Josephson and Rabi regimes of trapped Bose-Einstein condensates or atomic spin systems \cite{Smerzi,Leggett-2001,Zibold2010},  or more generally, to critical phenomena and quantum phase transitions in the Lipkin-Meshkov-Glick model \cite{Lipkin,Gilmore,Cejnar-2006,Vidal2006,Castanos,Ribeiro2007,Caprio,OKD2014,LMG-TO}. When damping is included, other effects can occur; in particular, highly anisotropic damping can be used to simulate the ``quantum Zeno effect'' \cite{Misra,Peres,Gonzalo}. 

As stressed by Feynman in his lectures, ``the same equations have the same solutions'' \cite{Feynman-lecture} which makes it possible to transfer intuition between different areas of physics. Although, certainly, not all quantum effects can be emulated by a classical pendulum, we believe that our approach can offer a different and fresh perspective on well known phenomena. Thus, physicists from the atomic and optical community may be gratified to observe the same effects on their next visit to a science museum. Some of those effects can even get a rather simple interpretation. 
Such deep connections of the pendulum motion to other seemingly unrelated and distinct areas of physics could be also useful for those who build and study Foucault pendulums, when they encounter certain strange and unexpected dynamical behavior.
%It might also help those who build  Foucault pendulums when they eucounter some unexpected behavior to a ppreciate the connection of their ``strange'' motions to deep physical problems studied elsewhere.

The paper is organized as follows. In Sec. \ref{Sec-Elliptic} we introduce the Stokes parameters and their relation to the pendulum motion. In Sec. \ref{Sec-Evolution} conservative evolution equations   are given and in Sec. \ref{Sec-Integrals} their integrals of motion are studied and conditions for stationary points and their stability are determined. Section \ref{Sec-Damping} deals with the damping. In Sec. \ref{Sec-Analogies} we study the analogies of the system to other areas of physics.  
We summarize our main conclusions in Sec. \ref{Sec-Conclusion}.

\section{Elliptical motion and the Stokes parameters}
\label{Sec-Elliptic}
%%%%%%%%%%%%%%%%%%%%%%%%%%%%%%%%%%%%%%%%%%%%%%%%%%%%%%%%%%%%%%%%%%%%%%%%%%%%%%

\begin{figure}%[h!]
\includegraphics[scale=0.45]{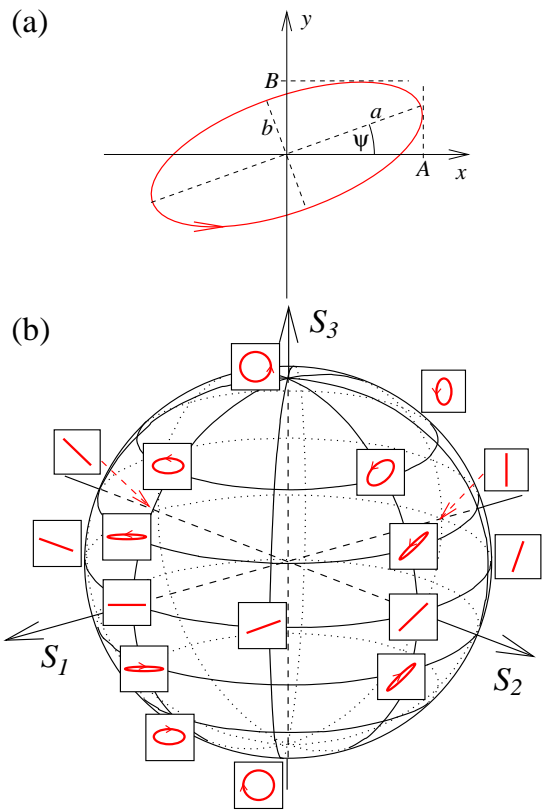}
\caption{\label{f-Poincare1}
(a) Scheme of the elliptical motion displaying the meaning of the parameters $a$, $b$, $\psi$, $A$ and $B$.
(b) Poincar\'{e} sphere with the corresponding  trajectories of the pendulum.
Circularly polarized trajectories occupy the poles, linearly polarized trajectories are localized on the equator and elliptical trajectories are on the rest of the surface.
}
\end{figure}

Let us assume that during a sufficiently short time interval  the horizontal projection of the pendulum motion can be approximated by an ellipse with semi-axes $a$ and $b$, and the major axis inclined at angle $\psi$ to the $x$-axis, as shown in Fig. \ref{f-Poincare1}(a).
The Cartesian variables are
\begin{eqnarray}
\label{xpsi}
x &=& a \cos \psi \cos \omega t - b \sin \psi \sin \omega t, \\
y &=& a \sin \psi \cos \omega t + b \cos \psi \sin \omega t.
\label{ypsi}
\end{eqnarray}
%\ref{f-elipsa1}.
The same ellipse can be described by  amplitudes $A$ and $B$ in the directions $x$ and $y$ and their phase difference $\phi$ as 
\begin{eqnarray}
\label{xA}
x &=& A \cos \omega t, \\
y &=& B \cos (\omega t - \phi),
\label{yB}
\end{eqnarray}
ignoring an unimportant constant shift 
of the temporal origin in going from Eqs. (\ref{xpsi})--(\ref{ypsi}) to  Eqs.  (\ref{xA})--(\ref{yB}).

In analogy to describing a general elliptical polarization of light, we introduce (dimensionless) Stokes parameters as 
\begin{eqnarray}
\label{Stokes30}
S_0 &\equiv& \frac{2}{L^2}\left(\langle x^2\rangle + \langle y^2 \rangle \right), \\
S_1 &\equiv& \frac{2}{L^2}\left(\langle x^2\rangle - \langle y^2 \rangle \right), \\
S_2 &\equiv& \frac{4}{L^2}\langle xy \rangle, \\
S_3 &\equiv& \pm \frac{4}{L^2}\sqrt{\langle x^2\rangle \langle y^2 \rangle - \langle xy \rangle ^2}, 
\label{Stokes3}
\end{eqnarray}
where $L$ is the pendulum length.
The average is over one period $T= 2\pi/\omega$, i.e., $\langle f(x,y)\rangle \equiv T^{-1}\int_{0}^{T}f(x,y) dt$, where $\omega = \sqrt{g/L}$ is the frequency of small oscillations of the pendulum and $g$ is the  acceleration due to gravity.
The sign in Eq. (\ref{Stokes3}) is positive (negative) for counterclockwise (clockwise) motion, respectively. The Stokes parameters satisfy
\begin{eqnarray}
S_0^2=S_1^2 + S_2^2+S_3^2 ,
\label{S02}
\end{eqnarray}
and they can be used to visualize the polarization state as a point with coordinates $(S_1,S_2,S_3)$ on a Poincar\'{e} sphere of radius $S_0$, as in Fig. \ref{f-Poincare1}(b). Note that although working with a different approach, the recently introduced concept of ``anisosphere'' \cite{Verreault-2017}  describing motion of the Foucault pendulum is equivalent to the Poincar\'{e} sphere.

%%%%%%%%%%%%%%%%%  F I G U R E %%%%%%%%%%%%%%%%%%%%%%
%%%%%%%%%%%%%%%%%  E N D % F I G U R E %%%%%%%%%%%%%%%%%%%%%%

The Stokes parameters can be
expressed in terms of the parameters of the ellipse as
\begin{eqnarray}
S_0 &=& \frac{a^2+b^2}{L^2}, \\
\label{Stokes1a}
S_1 &=&  \frac{a^2-b^2}{L^2} \cos 2\psi, \\
\label{Stokes2a}
S_2 &=&  \frac{a^2-b^2}{L^2} \sin 2\psi, \\
S_3 &=&  \frac{2ab}{L^2},
\label{Stokes3a}
\end{eqnarray}
or, alternately, 
\begin{eqnarray}
\label{Stokes0b}
S_0 &=& \frac{A^2+B^2}{L^2}, \\
\label{Stokes1b}
S_1 &=&  \frac{A^2-B^2}{L^2}, \\
S_2 &=&  \frac{2AB}{L^2} \cos \phi, \\
S_3 &=&  \frac{2AB}{L^2} \sin \phi.
\label{Stokes3b}
\end{eqnarray}
Parameters $S_0$ and $S_3$ have simple physical meanings: the energy of the pendulum is 
\begin{eqnarray}
E = \frac{1}{2}m\omega^2 L^2 S_0,
\end{eqnarray}
and the vertical component of its angular momentum is 
\begin{eqnarray}
{\cal L} = \frac{1}{2}m\omega L^2 S_3, 
\end{eqnarray}
where $m$ is the mass of the bob.

\section{Conservative time evolution}
\label{Sec-Evolution}
%%%%%%%%%%%%%%%%%%%%%%%%%%%%%%%%%%%%%%%%%%%%%%%%%%%%%%%%%%%%%%%%%%%%%%%%%%%%%%%
Let us first discuss three special kinds of evolution of the pendulum that will later be combined into unified motion.

\subsection{Anisotropic oscillations}
%%%%%%%%%%%%%
Assume that the frequency in $x$-direction is higher by $\Delta \omega$ than that in the $y$-direction, with $\Delta \omega \ll \omega$. 
This is typically caused by the
directional dependence of elasticity in the suspension mechanism,
or by asymmetric mass distribution in the physical pendulum as studied by Kamerlingh Onnes
\cite{Onnes}.
Then in Eqs. (\ref{xA}) and (\ref{yB}) the parameter $\phi$ varies with $\dot{\phi}=\Delta \omega$, whereas $A$ and $B$ are constant. Taking time derivatives of Eqs. (\ref{Stokes1b})---(\ref{Stokes3b}) one  finds 
\begin{eqnarray}
\label{StokesR1b}
\dot{S}_1 &=& 0, \\
\dot{S}_2 &=& -\Delta \omega S_3, \\
\dot{S}_3 &=&  \Delta \omega S_2,
\label{StokesR3b}
\end{eqnarray}
which corresponds to the rotation of the Poincar\'{e} sphere around $S_1$ as in Fig. \ref{f-Poincare2}(a).

\subsection{Motion in a rotating frame}
%%%%%%%%%%%%%%%%%
Let us assume that the frame of reference rotates around $z$ with angular velocity $\Omega$, as in the case of Foucault pendulum. The Coriolis force then causes rotation of the pendulum orbit with angular velocity $-\Omega$, i.e., the inclination angle $\psi$ of Eqs. (\ref{xpsi}) and (\ref{ypsi}) change at rate $\dot{\psi}=-\Omega$ while $a$ and $b$ remain constant. 
Taking time derivatives of Eqs. (\ref{Stokes1a})---(\ref{Stokes3a}) one  finds 
\begin{eqnarray}
\label{StokesR1a}
\dot{S}_1 &=& 2\Omega S_2, \\
\dot{S}_2 &=& -2\Omega S_1, \\
\dot{S}_3 &=&  0,
\label{StokesR3a}
\end{eqnarray}
which corresponds to the rotation of the Poincar\'{e} sphere around $S_3$ with {\em twice} the angular velocity of the coordinate system rotation (see Fig. \ref{f-Poincare2}(b)).

\subsection{Precession due to pendulum nonlinearity}
%%%%%%%%%%%%%%%%%%%%%%%%
As first derived by Airy \cite{Airy}, the elliptical trajectory of a pendulum precesses with angular velocity $\tilde \Omega$ proportional to the area of the ellipse 
\begin{eqnarray}
\tilde{\Omega} = \frac{3ab}{8L^2}\omega ,
\end{eqnarray}
the precession being in the same sense as the motion of the pendulum along the ellipse. This means that the inclination angle $\psi$ evolves as
\begin{eqnarray}
\dot{\psi} = \tilde{\Omega} = \frac{3}{16}\omega S_3.
\end{eqnarray}
Using this in the time derivatives of Eqs. (\ref{Stokes1a})---(\ref{Stokes3a}) yields
\begin{eqnarray}
\label{StokesT1b}
\dot{S}_1 &=& -\frac{3}{8}\omega S_2 S_3, \\
\dot{S}_2 &=& \frac{3}{8}\omega S_1 S_3, \\
\dot{S}_3 &=&  0.
\label{StokesT3b}
\end{eqnarray}
This corresponds to {\em twisting} the Poincar\'{e} sphere: Individual points rotate around the $S_3$-axis,  their angular velocity being proportional to the value of $S_3$. The corresponding motion is shown in Fig. \ref{f-Poincare2}(c).

\subsection{Combined motion}
\begin{figure}%[h!]
\centerline{\includegraphics[scale=0.25]{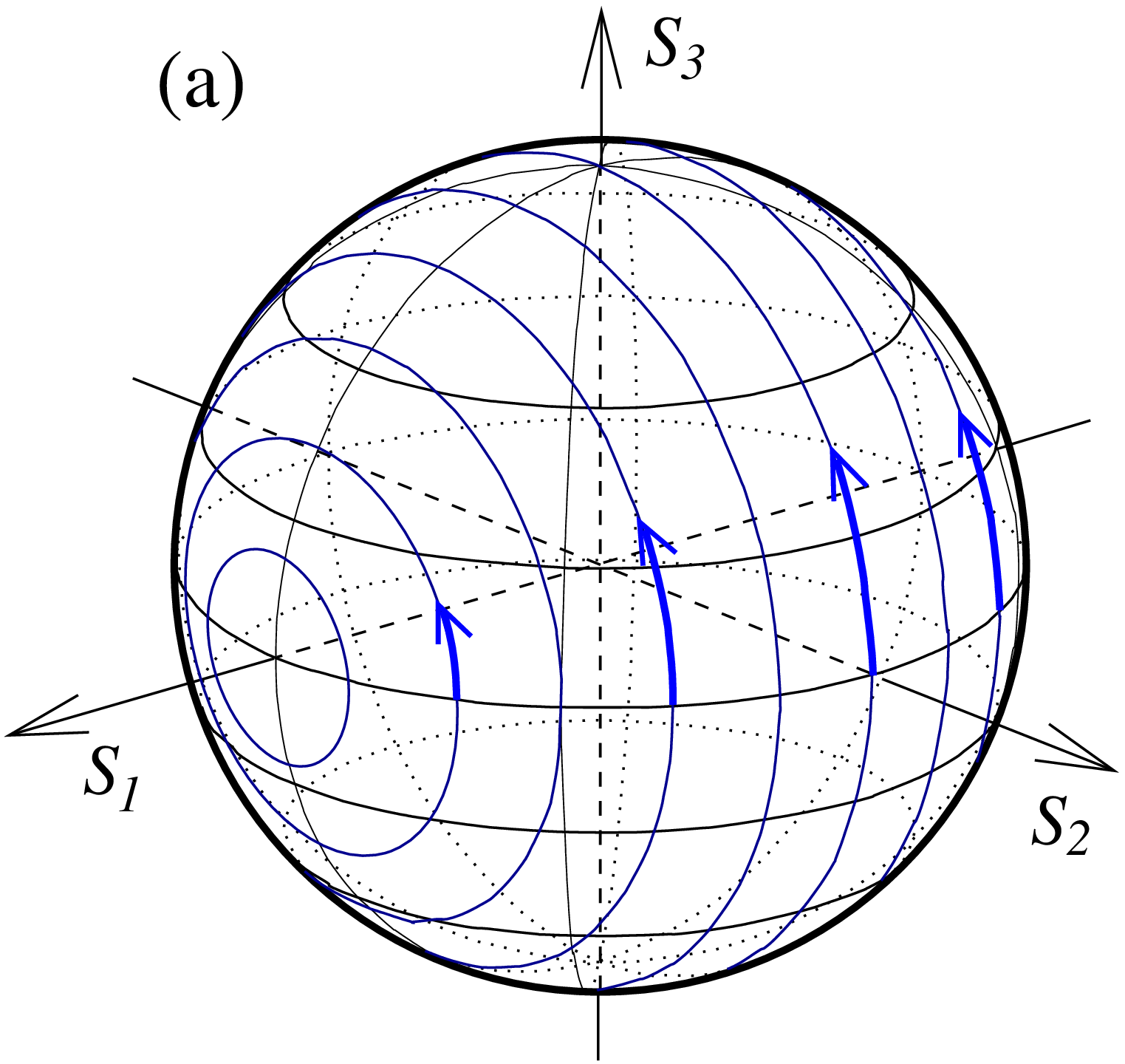}
\includegraphics[scale=0.25]{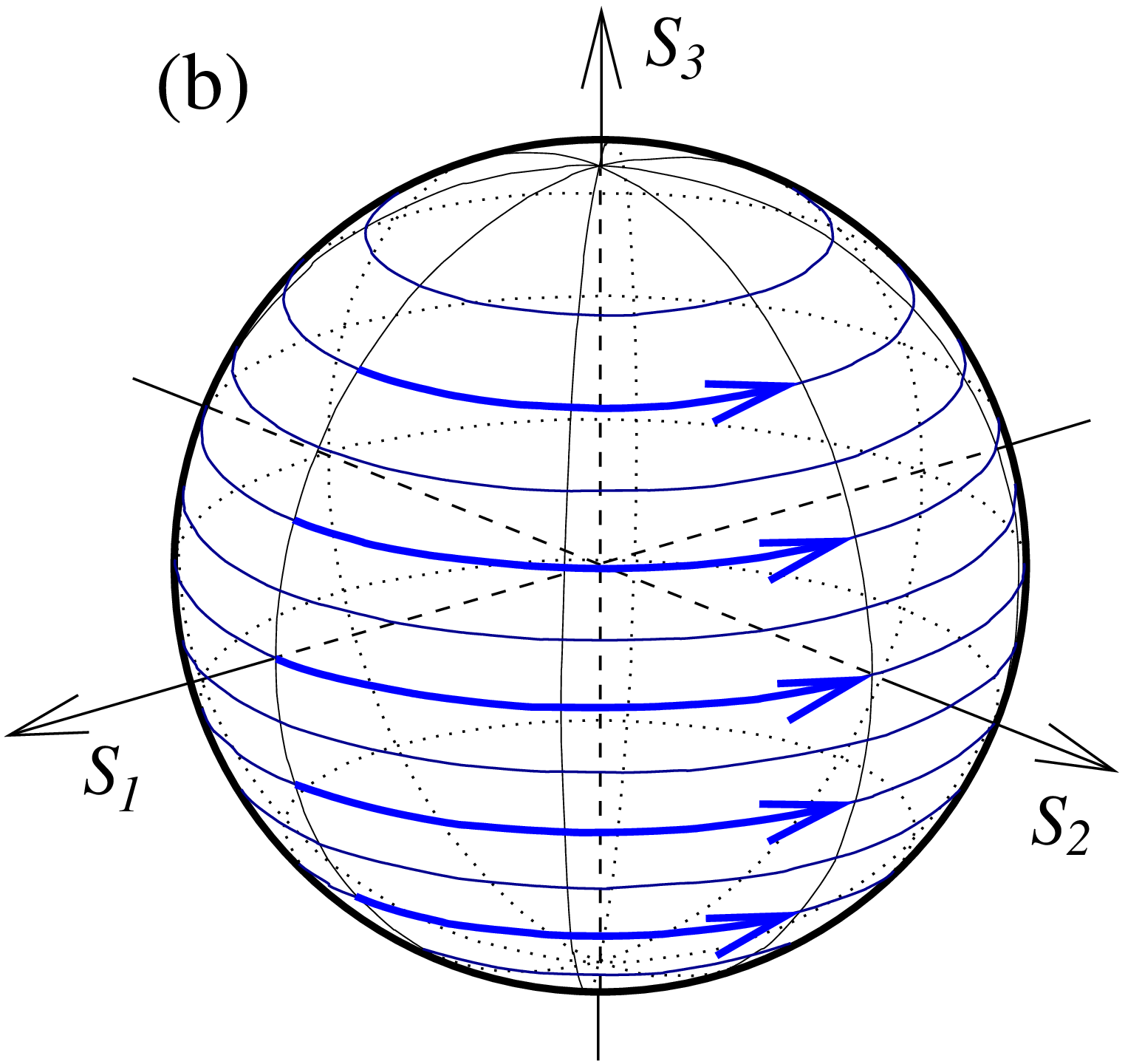}}
\centerline{\includegraphics[scale=0.25]{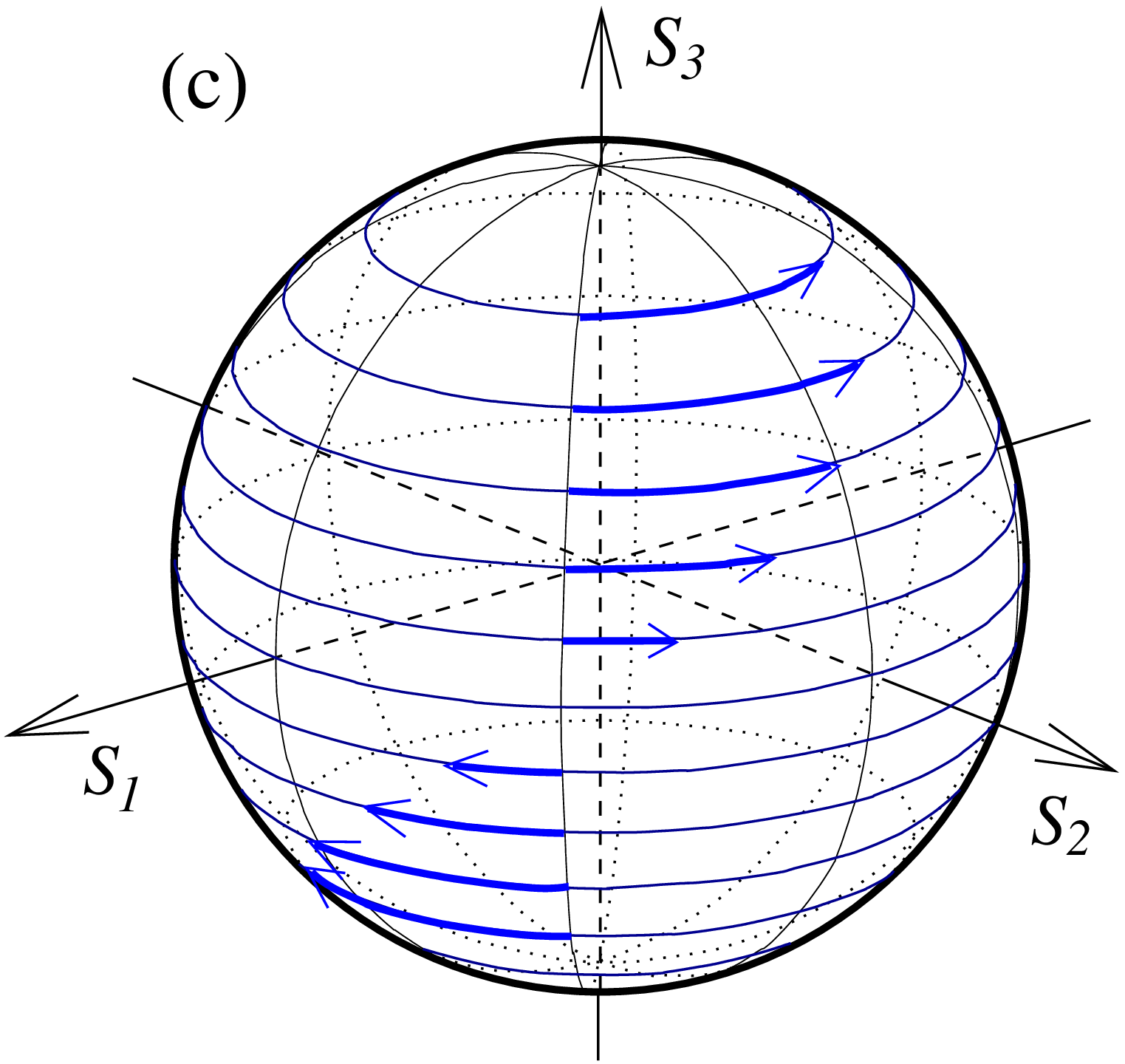}
\includegraphics[scale=0.25]{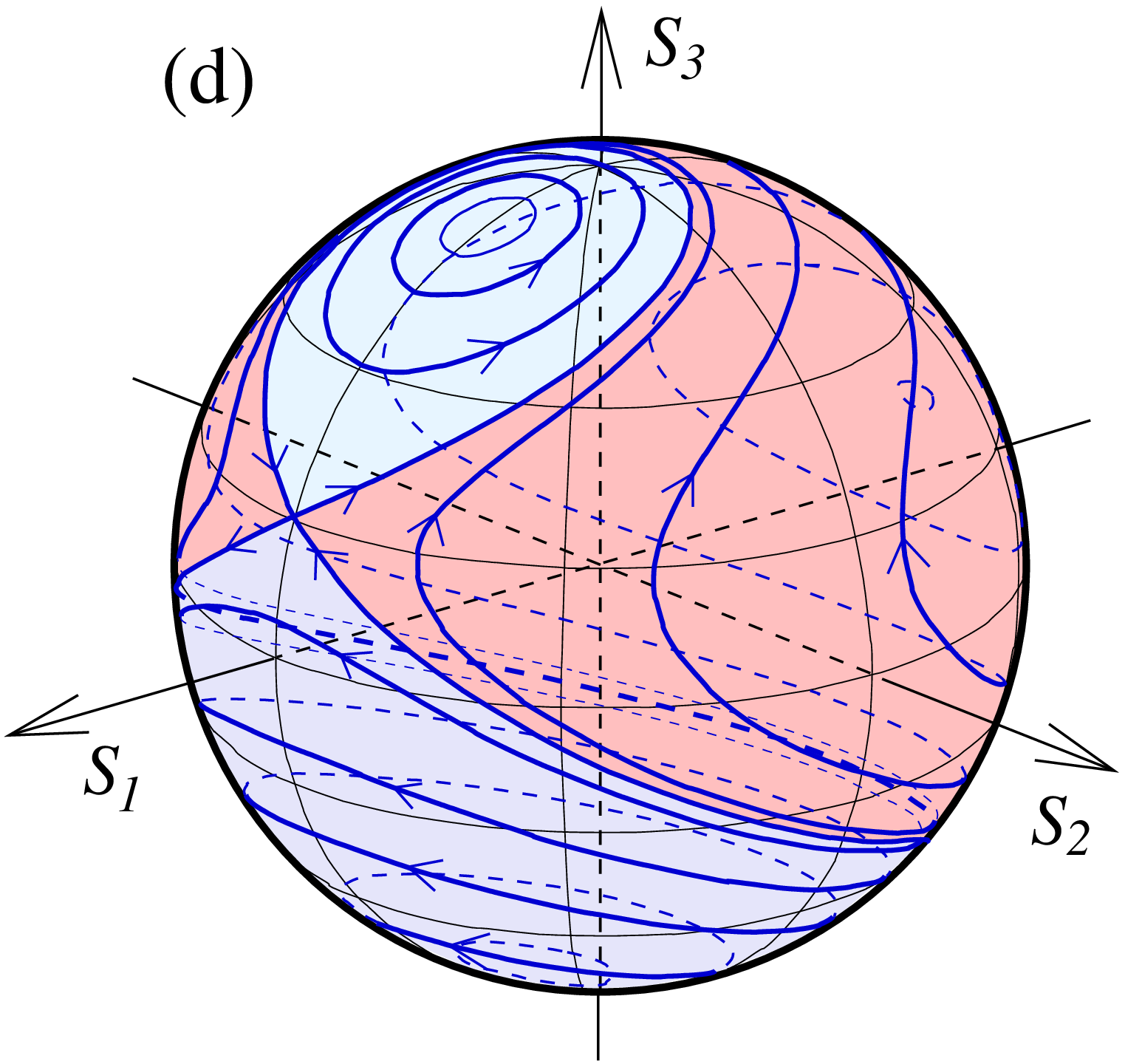}}
\caption{\label{f-Poincare2}
Poincar\'{e} sphere trajectories for an asymmetric harmonic oscillator (a), symmetric harmonic oscillator in a rotating frame (b), spherical pendulum with Airy precession (c), and asymmetric pendulum in a rotating frame (d).
}
\end{figure}

%%%%%%%%%%%%%%%%%%%%
Combining all the three effects, the resulting equations of motion are
\begin{eqnarray}
\begin{array}{cccccc}
\dot{S}_1 &=& & 2\Omega S_2 & & -\frac{3}{8}\omega S_2 S_3, \\
\dot{S}_2 &=& -2\Omega S_1 & & -\Delta \omega S_3 & + \frac{3}{8}\omega S_1 S_3, \\
\dot{S}_3 &=&  & \Delta \omega S_2.& &
\end{array}
\label{StokesComb}
\end{eqnarray}
This is a nonlinear set of equations that can be solved numerically, 
an example of which is shown  in Fig. \ref{f-Poincare2}(d). However, one can find integrals of motion from which trajectories can be expressed analytically.

%%%%%%%%%%%%%%%%%  F I G U R E %%%%%%%%%%%%%%%%%%%%%%
%%%%%%%%%%%%%%%%%  E N D % F I G U R E %%%%%%%%%%%%%%%%%%%%%%
% prvni verze Poincare-gen (d):  CHIx=0.5; CHIy=0; CHIz=1; OMEGAX=0; OMEGAY=0; OMEGAZ=1;
% Poincare-gen2: CHIx=0; CHIy=0; CHIz=1; OMEGAX=0; OMEGAY=0.5;  OMEGAZ=0.5;

\section{Integrals of motion, trajectories and stationary points}
\label{Sec-Integrals}
\begin{figure}%[h!]
\centerline{\includegraphics[scale=0.36]{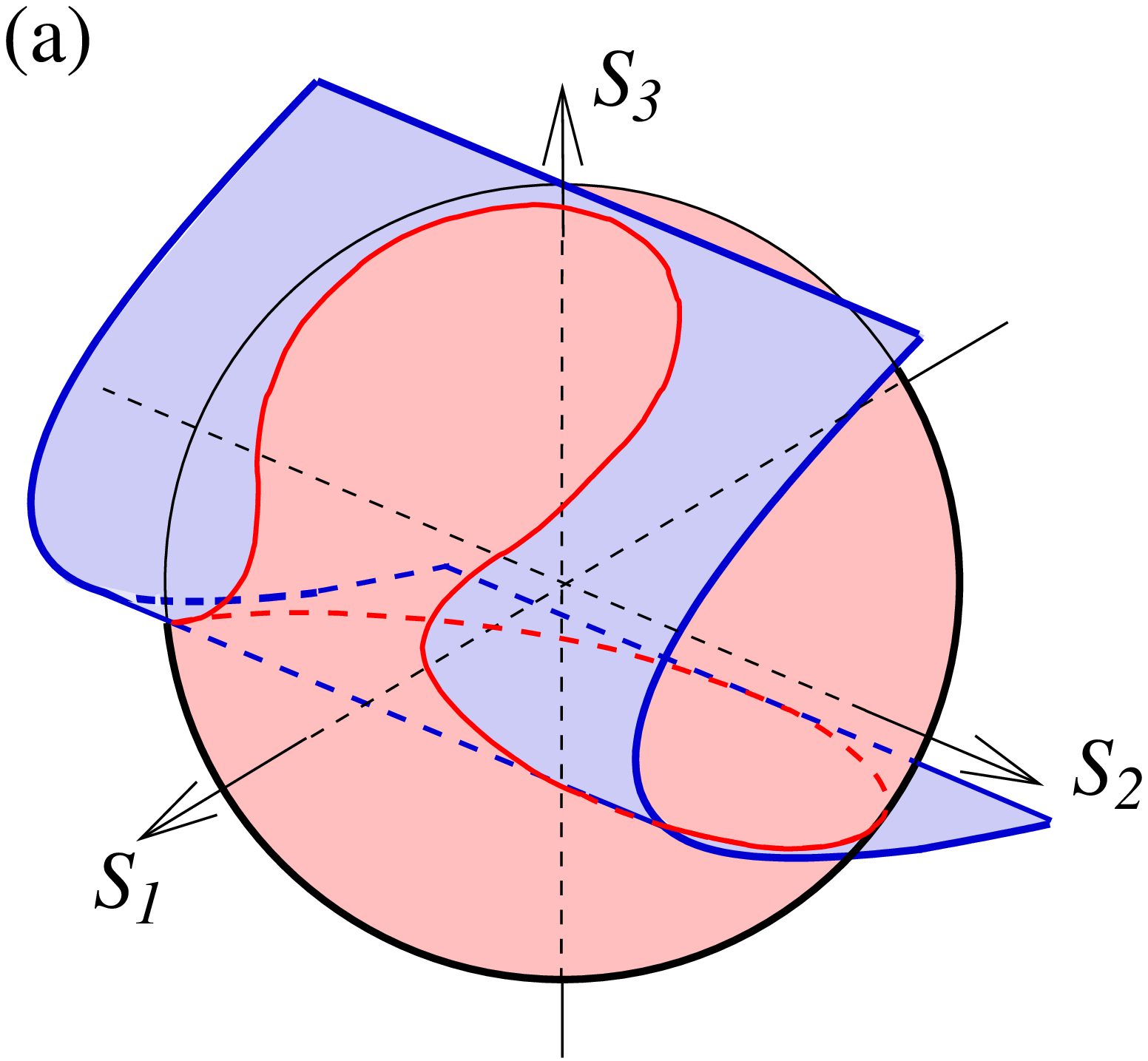}}
\centerline{\includegraphics[scale=0.46]{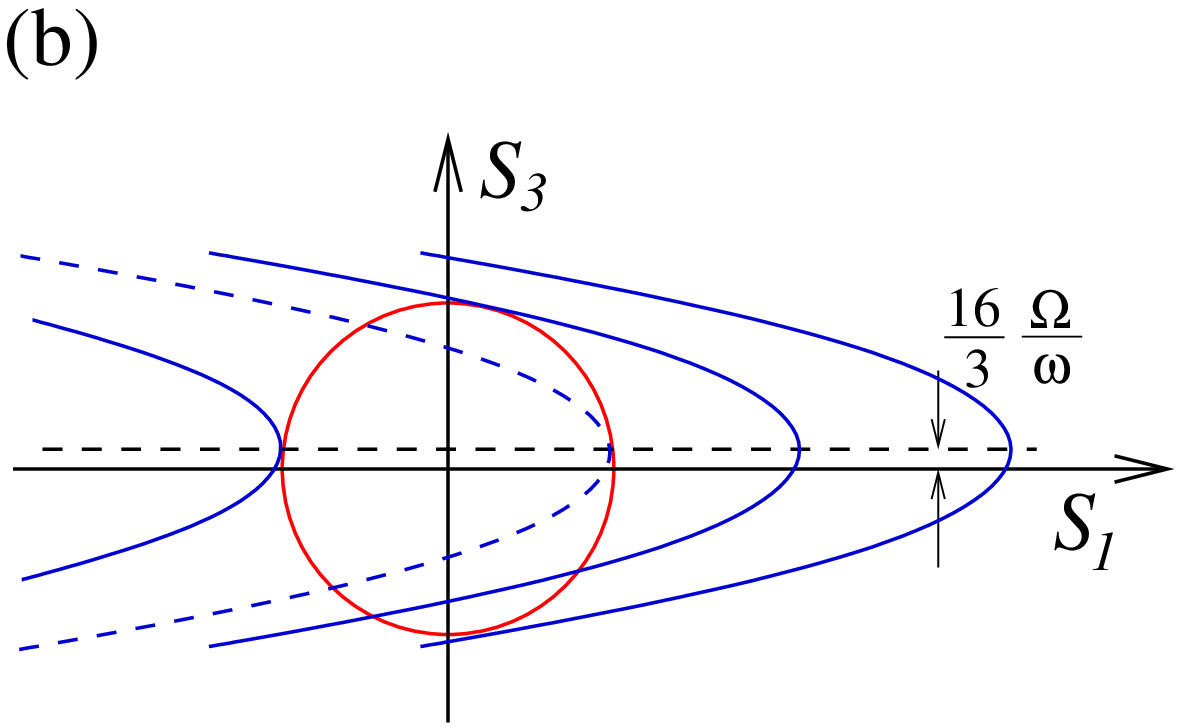}}
%\centerline{\epsfig{file=fintersect2,width=0.5\linewidth}
%\epsfig{file=f-paraboly,width=0.5\linewidth}}
\caption{\label{f-Poincare3}
(a) Poincar\'{e} sphere trajectory as an intersection of the sphere $S_0=$ const and a parabolic cone $H=$ const. (b) Stationary points as points of touch of the sphere and parabolic cones. The full lines represent the cones touching at the stable stationary points, the broken line represents the cone touching at the unstable stationary point.
}
\end{figure}

%%%%%%%%%%%%%%%%%%%%%%%%%%%%%%%%%%%%%%%%%%%%%%%%%%%%%%%%%%%%%%%%%%%%%
There are two integrals of motion related to the set (\ref{StokesComb}), namely $S^2_0$ of Eq. (\ref{S02}) and $H$ given by
\begin{eqnarray}
H = \Delta \omega S_1 - 2 \Omega S_3 + \frac{3}{16}\omega S_3^2 .
\label{eqH}
\end{eqnarray}
This can be checked by taking time derivatives $dS_0^2/dt$ of  Eq. (\ref{S02}) and $d{H}/dt$ of Eq. (\ref{eqH}) and applying there Eqs. (\ref{StokesComb}).
Therefore, trajectories in the $(S_1,S_2,S_3)$ coordinates are lines of intersection of a sphere $S_0^2 =$ const and a parabolic cylinder $H = $ const (see Fig. \ref{f-Poincare3}(a)). 

%%%%%%%%%%%%%%%%%  F I G U R E %%%%%%%%%%%%%%%%%%%%%%
%%%%%%%%%%%%%%%%%  E N D % F I G U R E %%%%%%%%%%%%%%%%%%%%%%

The trajectories reduce to stationary points where the sphere barely touches the parabolic cylinder (see Fig. \ref{f-Poincare3}(b)). They can be found from 
Eq. (\ref{StokesComb}) by setting the left-hand side equal to zero. 
%"Thus for delta omega != 0 one naturally finds S_2 =  0, since the first derivative of S_3  and S_1 (31) is then simply also equal to zero, as well as the first derivative of S2. By following expression of S_3 and inserting it into equation S_1 ** 2 + S_3 ** 2 = S_0 ** 2 one gets..."
Thus for $\Delta \omega \neq 0$ one finds $S_2=0$ and 
by expressing $S_3$ and inserting it into equation $S_1^2+S_3^3=S_0^2$ one gets
\begin{eqnarray}
S_1^4 - \frac{8\Delta \omega}{3\omega} S_1^3 + \left[ \left(\frac{4\Omega}{3\omega}\right)^2 
+  \left(\frac{4\Delta \omega}{3\omega}\right)^2 -S_0^2\right]S_1^2 
\nonumber \\
+ \frac{8\Delta \omega}{3 \omega}S_0^2 S_1 -  \left(\frac{4\Delta \omega}{3\omega}\right)^2 S_0^2 = 0.
\label{Eq4th}
\end{eqnarray}
Eq. (\ref{Eq4th}) has up to 4 real roots for $S_1$ in the interval between $-S_0$ and $S_0$.

In the special case of a symmetric pendulum, $\Delta \omega=0$, the parabolic cylinder reduces to a pair of planes perpendicular to $S_3$. Two stable stationary points are the poles of the Poincar\'{e} sphere $S_1=S_2=0$, $S_3=\pm S_0$ corresponding to circular polarizations. In case of $S_0>16 |\Omega|/(3\omega)$ there is a circle of unstable stationary points with $S_3 = 16 \Omega/(3\omega)$.

Another special case corresponds to an asymmetric pendulum with no rotation, $\Omega = 0$. Here the parabolic cylinder has a plane of symmetry $S_1=0$ and two of the stationary points are $S_1=\pm S_0$, $S_2=S_3=0$. If $S_0> 8|\Delta \omega|/(3\omega)$, then the radius of the sphere is bigger than one of the principal radii of the cylinder at the point of their contact. In this case, the point where the
cylinder touches the sphere from inside is unstable.
This can be seen by observing nearby trajectories that occur if the cylinder is shifted by a slight change of $H$: locally the trajectories are hyperbolas rather than ellipses as would be the case of a stable stationary point. Apart from the stable and unstable points at  $S_1=\pm S_0$,
  two additional stable stationary points occur at $S_1=8\Delta \omega/(3\omega)$, $S_2=0$, $S_3=\pm \sqrt{S_0^2-S_1^2}$.

%%%%%%%%%%%%%%%%%  F I G U R E %%%%%%%%%%%%%%%%%%%%%%
\begin{figure}%[h!]
\centerline{\includegraphics[scale=0.4]{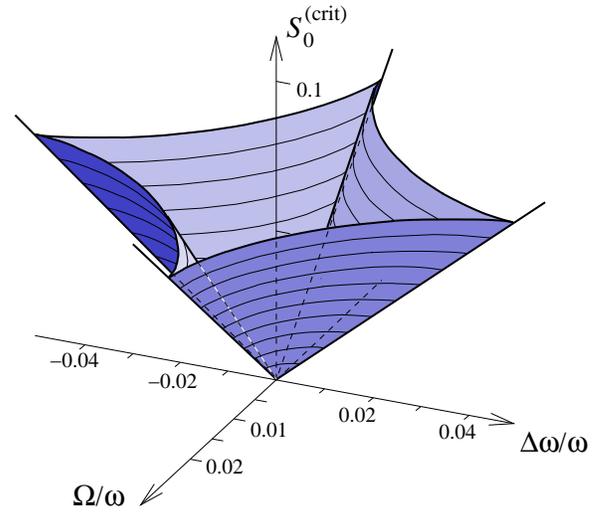}}
%\centerline{\epsfig{file=f-crit,width=0.6\linewidth}}
\caption{\label{f-crit}
Dependence of the critical value $S_0^{\rm (crit)}$ on $\Delta \omega$ and $\Omega$ as in Eq. (\ref{S0crit}). Inside the surface the nonlinearity is dominant and unstable stationary points exist; outside the linear terms are dominant and the equations of motion have just two stable stationary points.
}
\end{figure}
%%%%%%%%%%%%%%%%%  E N D % F I G U R E %%%%%%%%%%%%%%%%%%%%%%

In the general case of four stationary points, three of them are stable and one is unstable. Again, the unstable stationary point corresponds to the parabolic cylinder touching the
sphere  from inside at a point where one of the principal radii of the cylinder is smaller than the radius of the sphere.  From a simple geometric consideration we find the critical value of the radius
\begin{eqnarray}
\label{S0crit}
S_0^{\rm (crit)} = \frac{8}{3}\frac{\left[\Delta \omega^{2/3} + (2\Omega)^{2/3} \right]^{3/2}}{\omega}
\end{eqnarray}
above which such kind of contact can occur (see \ref{Sec-Ap1} for the derivation). This function is shown in Fig. \ref{f-crit}.
Thus, for $S_0>S_0^{\rm (crit)}$ there are three stable and one unstable stationary points. For  $S_0=S_0^{\rm (crit)}$ two stable points merge with the unstable point forming one stable point, and for  $S_0<S_0^{\rm (crit)}$ only two stable points exist.

In the case of two stationary points all trajectories encircle the stationary points in the same sense. In the case of four stationary points the Poincar\'{e} sphere is divided into three separate areas, each containing one stationary point encircled by the trajectories. These areas shown in different colors in Fig. \ref{f-Poincare2}(d) are separated by a line called {\em separatrix} that crosses itself at the unstable stationary point.

In terms of the pendulum motion, consider first a nonrotating system with $\Omega=0$, $\Delta \omega \neq 0$, and $S_0>S_0^{\rm (crit)}$. Two of the stationary points correspond to elliptical polarizations with the larger axis along the fast direction of the pendulum, the pendulum orbiting in opposite sense in the two cases. The third stationary point corresponds to the linear polarization along the slow direction of the pendulum. Linear polarization along the fast direction of the pendulum is unstable; if the pendulum is swung in this direction, the motion would soon change into elliptical and evolve depending on the exact value of the initial condition. 
In rotating systems with $\Omega\neq 0$ the stationary points move towards the north or south pole, depending on the sign of $\Omega$. Each of the stationary points now corresponds to some general elliptical polarization.

\section{Damped motion}
\label{Sec-Damping}
%%%%%%%%%%%%%%%%%%%%%%%%%%%%%%%%%%%%%%%%%%%%%%%%%%%%%%%%%%%%%%%%%%%%%%%%%%%%%
Various kinds of damping can be included in the model to modify the equations of motion. Here we study
a particularly simple situation of linear, in general anisotropic damping. Assume a damping force  $\vec{F}\equiv (F_x,F_y) = -(c_x \dot{x},c_y \dot{y})$ where $c_{x,y} \ge 0$ are constants. The special case of $c_x=c_y$ corresponds, e.g., to viscous damping of the pendulum bob due to air drag with low Reynolds numbers. The anisotropic case of $c_x \neq c_y$ could stem, e.g., from viscous damping in a  Cardan suspension of the pendulum. Using these forces in the equations of weakly damped harmonic oscillators one finds the time change of the amplitudes as
\begin{eqnarray}
\dot{A} &=& -\frac{\gamma_x}{2} A, \\
\dot{B} &=& -\frac{\gamma_y}{2} B,
\end{eqnarray}
where $\gamma_{x,y} = c_{x,y}/m$, which leads to
\begin{eqnarray}
\dot{S}_0 &=& -\frac{\gamma_x + \gamma_y}{2} S_0 - \frac{\gamma_x-\gamma_y}{2} S_1, \\
\label{DampS1}
\dot{S}_1 &=& -\frac{\gamma_x - \gamma_y}{2} S_0 - \frac{\gamma_x+\gamma_y}{2} S_1, \\
\dot{S}_2 &=&  - \frac{\gamma_x+\gamma_y}{2} S_2, \\
\dot{S}_3 &=&  - \frac{\gamma_x+\gamma_y}{2} S_3.
\end{eqnarray}
Note that in Eq. (\ref{DampS1}) $S_0$ can be expressed as $S_0=\sqrt{S_1^2+S_2^2+S_3^2}$ so that three equations (although nonlinear) for $S_{1,2,3}$ are enough to govern the Stokes vector motion.

In the isotropic case of $\gamma_x=\gamma_y \equiv \gamma$, all the Stokes parameters are damped with the same rate, $\dot{S}_k=-\gamma S_k$, for $k=0,\dots ,3$. Another extreme case is the unidirectional damping with  $\gamma_x\equiv \gamma$, $\gamma_y=0$, where we get
\begin{eqnarray}
\dot{S}_0 &=& -\frac{\gamma}{2} S_0 - \frac{\gamma}{2} S_1, \\
\dot{S}_1 &=& -\frac{\gamma}{2} S_0 - \frac{\gamma}{2} S_1, \\
\dot{S}_2 &=&  - \frac{\gamma}{2} S_2, \\
\dot{S}_3 &=&  - \frac{\gamma}{2} S_3.
\end{eqnarray}
In this case the maximum dissipation rate occurs for $S_{2,3}=0$, $S_1=S_0$, where $\dot{S}_{0,1}=-\gamma S_{0,1}$, whereas no dissipation takes place for  $S_{2,3}=0$, $S_1=-S_0$, where $\dot{S}_{0,1}=0$.

In a similar way one can also treat more general forms of damping, although the equations become more involved and will be studied elsewhere.

\section{Relation to other areas of physics}
\label{Sec-Analogies}
%%%%%%%%%%%%%%%%%%%%%%%%%%%%%%%%%%%%%%%%%%%%%%%%%%%%%%%%%%%%%%%%%%%%%%%%%%%%%
\subsection{Polarized light and Zeno effect}
%%%%%%%%%%%%%%%%%%%
The poles of the Poincar\'{e} sphere correspond to circularly polarized light whereas the points on the equator represent different linearly polarized states. Rotation around the vertical axis as in Fig. \ref{f-Poincare2}(b) corresponds to rotation of the polarization plane by optically active media such as a sugar solution. Rotation around a horizontal axis as in  Fig. \ref{f-Poincare2}(a) is caused by anisotropic media in which vertically polarized light propagates with different speed from horizontally polarized light. Devices made of such materials are used, for example, to convert circularly polarized light into linear (a quater-wave plate) such as in some glasses used for 3D cinema. 

Anisotropic damping corresponds to dichroism, i.e., to light absorption depending on the linear polarization. A polarization filter can be modeled by extremely anisotropic damping, e.g., a pendulum with a Cardan suspension with vanishing friction in one direction and substantial friction in the perpendicular direction: after a certain time, only the oscillations in the undamped direction survive.

An optical ``Zeno effect'' \cite{Misra} can be observed when  a sequence of polarizers is inserted into the optically active medium \cite{Peres,Gonzalo}. 
Assume first a slab of optically active medium of such a width that the linear polarization of a passing light beam rotates by $\pi/2$. If two filters of the same polarization direction are added, one at the input and the other at the output of the slab, then no light can go through. Insert now a sequence of equidistant filters into the medium, their polarization directions being the same as that of the input and output filters. Since each of the filters projects the beam polarization to that of the input filter, some light can now pass through. 
In the limit of infinitely many perfect filters the system becomes perfectly transparent for the light that passes through the first filter.
This is a seeming paradox, since adding absorbing elements actually decreases the resulting absorption. 
In quantum physics the effect corresponds to  measurements that project a system to the input state. 
Frequent measurements then lead to the inhibition of evolution of the system: One can ``stop the motion by watching''.

An analogous effect can be observed with a Foucault pendulum and highly anisotropic damping. With no damping,  the plane of swinging would rotate by $\pi/2$ in  time $T_{\pi/2}=\pi/(2\Omega)$. 
With anisotropic damping $0\approx \gamma_y < \Omega \ll \gamma_x,$ and with the pendulum initially swinging in the undamped direction $y$, it will stay swinging in the same direction with almost no losses. If, on the other hand, the damping is switched on only at the beginning  and after time  $T_{\pi/2}$ (equivalent to only the input and output filters present), then all the pendulum's energy will be absorbed by the damping mechanism. 

%There are no common optical devices implying the twisting transformation of  Fig. \ref{f-Poincare2}(c).

\subsection{Spin systems and Bose-Einstein condensates}
%%%%%%%%%%%%%%%%%
 Eq. (\ref{eqH}) can be written in an operator form to define a quantum Hamiltonian 
\begin{eqnarray}
\hat{H} = \Delta \omega \hat{S}_1 - 2 \Omega \hat{S}_3 + \frac{3}{16}\omega \hat{S}_3^2 ,
\label{eqHQuant}
\end{eqnarray}
 with $\hat{S}_{1,2,3}$  commuting as $[\hat{S}_k,\hat{S}_l]=i\epsilon_{jkl}\hat{S}_j$, where $\epsilon_{jkl}$ is the Levi-Civita symbol, and Einstein summation convention is used. Operators  $\hat{S}_{1,2,3}$ can be used to describe the collective spin of $N$ spin-half particles, with
$\hat{S}_0^2 \equiv \hat{S}_1^2+\hat{S}_2^2+\hat{S}_3^2$  being a constant related to the particle number as $\hat{S}_0^2 =\frac{N}{2}\left(\frac{N}{2}+1 \right)$. The evolution can be expressed in the form of Heisenberg equations $d\hat{S}_k/dt = i [\hat{H},\hat{S}_k]$ with $\hbar=1$. The resulting equations of motion correspond to Eq. (\ref{StokesComb}) with operator products being symmetrized as $S_k S_l \to \frac{1}{2}\left(\hat{S}_k \hat{S}_l + \hat{S}_l \hat{S}_k\right)$.

Points on the Poincar\'{e} sphere (in spin physics commonly called the Bloch sphere) correspond to spin-polarized states along the corresponding directions. Thus, for example, the north pole corresponds to all spins up whereas the south pole to all spins down. The dynamics described in Sec. \ref{Sec-Evolution} corresponds to several well-known effects of spin physics. If the spin-up and spin-down states have 
different energies then the sphere rotates around $S_3$  as in Fig.   \ref{f-Poincare2}(b) at a rate proportional to the energy difference. 
To rotate the sphere around a horizontal axis  as in Fig.   \ref{f-Poincare2}(a), a resonant microwave field can be applied. The spin then exhibits Rabi oscillations. The rotation rate is proportional to the intensity of the field and the orientation of the rotational axis depends on the phase of the field.

%%%%%%%%%%%%%%%%%  F I G U R E %%%%%%%%%%%%%%%%%%%%%%
\begin{figure}%[h!]
%\centerline{\epsfig{file=f-spectra1,width=0.5\linewidth}
%\epsfig{file=f-spectra2,width=0.5\linewidth}}
\centerline{\includegraphics[scale=0.32]{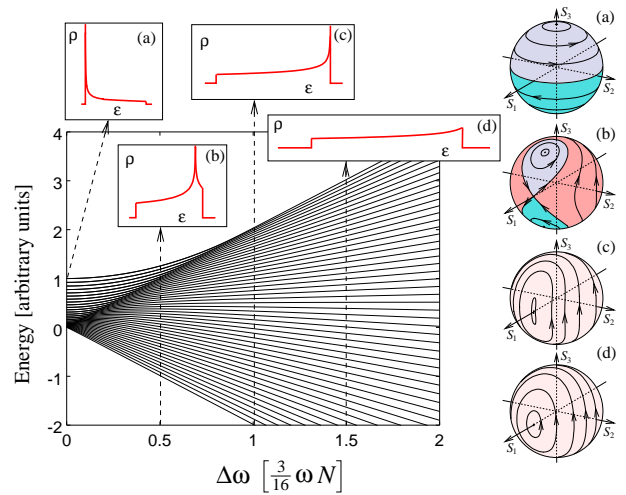}}
%\centerline{\epsfig{file=f-spectra1,width=0.9\linewidth}}
\caption{\label{f-Spectra1}
Dependence of eigenvalues of the Hamiltonian (\ref{eqHQuant}) on $\Delta \omega$ for $N=50$ and $\Omega=0$. The insets show the spectral densities $\rho(\epsilon)$ in the limit of $N\to \infty$, i.e., the relative numbers of eigenstates in rescaled energy intervals $d\epsilon$ for $\Delta \omega = 0$ (a), $\Delta \omega = 0.5 \cdot \frac{3}{16}\omega N$ (b),  $\Delta \omega =  \frac{3}{16}\omega N$ (c), and  $\Delta \omega = 1.5 \cdot \frac{3}{16}\omega N$ (d). The Poincar\'{e} spheres on the right show the corresponding classical pendulum trajectories 
%of Eqs. (\ref{StokesComb}) 
with substituting $S_0$ for $N/2$.
The regime of dominant nonlinearity is in the range of $\Delta \omega$ between cases (a) and (c) and is characterized by a diverging peak of the density of states that travels from the bottom of the energy spectrum (a) to the top (c). In the pendulum motion this corresponds to the existence of unstable stationary points. In case (a) the whole equator is formed from unstable stationary points, in the regime between (a) and (c), with a particular example of case (b) a single unstable point exists. Case (c) corresponds to the critical value of  $\Delta \omega$ where the unstable point merges with two stable points resulting in a stable stationary point. For higher $\Delta \omega$, as in the example (d), the density of states is smooth and there are only two stationary points in the pendulum motion that are stable.
}
\end{figure}
%%%%%%%%%%%%%%%%%  E N D % F I G U R E %%%%%%%%%%%%%%%%%%%%%%

The twisting operation as in Fig.   \ref{f-Poincare2}(c) is achieved in collective spin systems with mutual interactions. The corresponding  operation is used to generate {\em spin squeezing}, as first proposed by Kitagawa and Ueda \cite{Kitagawa} and experimentally demonstrated in \cite{Esteve2008,Gross2010,Riedel,Leroux-2010}. To understand it,
consider a collection of $N$ spins initially prepared each in the same superposition $2^{-1/2}(|\uparrow \rangle + |\downarrow \rangle)$. The collective state can be visualized, on the Bloch sphere of radius $N/2$, as a circle with its center at the equator and diameter (characterizing the uncertainty of spin projections) $\sim \sqrt{N}$. The twisting operation governed by Hamiltonian $\hat{H} \propto \hat{S}_3^2$ then deforms the circle, stretching it in one direction and squeezing in the perpendicular one. As a result, one obtains a collective spin state with decreased (``squeezed'') noise of some observable quantity. This quantity then can be used for enhanced precision measurements. Various combinations 
of the linear terms together with the quadratic one in Hamiltonian  (\ref{eqH}) can be used to optimize the squeezing procedure. In this way, the dynamics of Fig.   \ref{f-Poincare2}(d) corresponds to a variant of the ``twist-and-turn'' scenario for spin squeezing \cite{Opatrny2015,Muessel}. Imagine a state located at the unstable stationary point of  Fig.   \ref{f-Poincare2}(d). Its uncertainty circle is very efficiently stretched and squeezed in the directions of the arrows along the separatrix.

The same formalism can be used to describe dynamics of a Bose-Einstein condensate (BEC) located in a double-well trap  \cite{Smerzi} or in a ring trap with grating \cite{OKD2014}. Rather than two spin orientations, the atoms can have two different localizations of the double-well or two orbital states in the ring. Tunneling of atoms between the wells corresponds to the rotation of the sphere around a horizontal axis, and mutual repulsion (or attraction) of the atoms leads to the twisting operation. Circling around one of the stable stationary points near a pole corresponds to an effect called ``self-trapping'' of the BEC in a ``Josephson regime'' \cite{Leggett-2001}: even though a single atom would tunnel back and forth between the two wells, the atomic interaction can keep a BEC in a single well. An analogous effect can be seen in an asymmetric pendulum: in linear regime, if initialized with a circular orbit, the pendulum would exhibit ``Rabi flips'', switching between clockwise and counterclockwise orientations of the orbit. In the nonlinear regime (sufficiently high amplitude) the pendulum can be ``self-trapped'' in one of these orbital orientations.

\subsection{``Spin squeezing'' with a classical pendulum}
%%%%%%%%%%%%%%%%%%%%%%%%%%%%
Consider an ensemble of identical pendulums, each initialized  near the $x$-oriented linearly polarized state. To be more specific, in the initial ensemble, the mean values of the Stokes parameters are $\bar{S_1} \approx S_0$, $\bar{S_2}=0$ and $\bar{S_3}=0$, and the standard deviations are $\Delta S_2 = \Delta S_3 \equiv \Delta \ll S_0$, with no correlation between $S_2$ and $S_3$. After being left to evolve, the elliptical states with positive $S_3$ move by the Airy precession in the opposite direction compared to states with negative  $S_3$. As a result,  $S_2$ and $S_3$ become correlated and the ensemble forms a stretched area on the Poincar\'{e} sphere. After a sufficiently long time $\tau$ with $8/(3 \omega S_0) \ll \tau \lesssim 4/(3\omega \Delta)$ the ensemble area has approximate dimensions $\Delta_+ \approx (3/8)S_0 \omega \tau \Delta$ and $\Delta_- \approx 8\Delta/(3S_0 \omega \tau)$ and is tilted by angle $\alpha \approx 8/(3S_0\omega \tau)$ from the equator. When $\Delta_- < \Delta$ the ensemble is ``squeezed''. To remove the correlation and see the ``squeezing'' directly as in the spin-squeezing experiments, one can temporarily switch on the pendulum anisotropy to rotate the sphere around $S_1$ by angle $\alpha$ so as to align the stretched area with the equator. The ensemble is now squeezed in $S_3$, meaning that the elliptical component is suppressed and all the states are now closer to the linear polarization than at the beginning. The price is the increased uncertainty in the orientation of the linear polarization corresponding to the ensemble being stretched along the equator.

Note that squeezing with classical Hamiltonians has been explored in \cite{Opatrny-classqueez} with the focus on finding a general formula for the squeezing rate.

\subsection{Lipkin-Meshkov-Glick model and quantum phase transitions}
%%%%%%%%%%%%%%%%%%%%%%%%%%%%
In 1965 Lipkin, Meshkov and Glick (LMG) formulated a toy model of multiparticle interaction that can be, under certain conditions, solved exactly, and thus serve as a basis for testing various approximation methods \cite{Lipkin}. Although the original motivation was modeling energy spectra of atomic nuclei, the scheme turned out to be useful for studying critical phenomena in much more general systems. It has been shown that in the LMG model ground-state, quantum phase transition can occur \cite{Gilmore}, a concept later generalized to excited-state quantum phase transitions \cite{Cejnar-2006}.
Since then, phase transitions in the LMG model have been studied in great detail (see, e.g., \cite{Vidal2006,Castanos,Ribeiro2007,Caprio,OKD2014}). Recently the LMG phase transitions have been shown to correspond to transitions between regimes of classical motion of rigid bodies with rotors  \cite{LMG-TO}. 

%{Gilmore,Cejnar-2006,Castanos,Ribeiro2007,Caprio}.

%%%%%%%%%%%%%%%%%  F I G U R E %%%%%%%%%%%%%%%%%%%%%%
\begin{figure}%[h!]
%\centerline{\epsfig{file=f-spectra1,width=0.5\linewidth}
%\epsfig{file=f-spectra2,width=0.5\linewidth}}
%\centerline{\epsfig{file=f-spectra2,width=0.9\linewidth}}
\centerline{\includegraphics[scale=0.32]{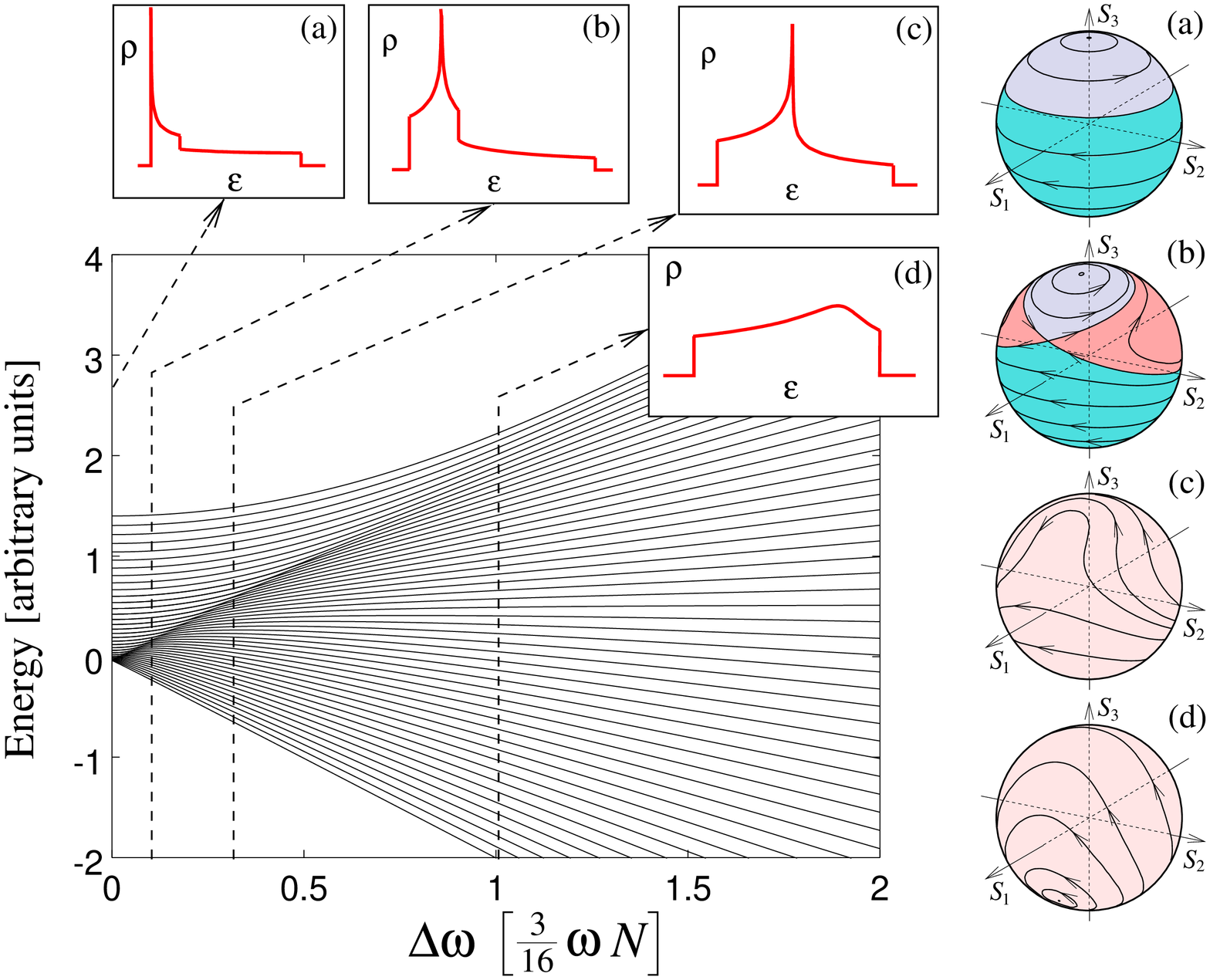}}
\caption{\label{f-Spectra2}
Same as Fig. \ref{f-Spectra1}, but with $\Omega =0.2 \cdot  \frac{3}{16}\omega N$. The insets and the Poincar\'{e} spheres correspond to  $\Delta \omega = 0$ (a), $\Delta \omega = 0.1 \cdot \frac{3}{16}\omega N$ (b),  $\Delta \omega =   0.309 \cdot \frac{3}{16}\omega N$ (c), and  $\Delta \omega = \frac{3}{16}\omega N$ (d). Case (c) corresponds to the critical value of $\Delta \omega$ where the unstable stationary point merges with one of the stable points. Differently from Fig.  \ref{f-Spectra1}, in the nonlinear regime the density of states has a discontinuity (apart from the diverging peak) that corresponds to a local maximum of the Hamiltonian. For the critical value of  $\Delta \omega$ the diverging peak merges with the discontinuity, forming a smooth maximum of $\rho$ for higher values of   $\Delta \omega$.  
}
\end{figure}
%%%%%%%%%%%%%%%%%  E N D % F I G U R E %%%%%%%%%%%%%%%%%%%%%%

The LMG Hamiltonian has the form
\begin{eqnarray}
\label{HLMG}
\hat{H}_{\rm LMG}=\epsilon  \hat{S}_3 + V(\hat{S}_1^2- \hat{S}_2^2) + W(\hat{S}_1^2+ \hat{S}_2^2),
\end{eqnarray}
where $\epsilon, V$ and $W$ are real parameters. In the special case of $V=0$ the LMG Hamiltonian is equivalent to 
\begin{eqnarray}
\hat{H}=\epsilon \hat{S}_3 - W \hat{S}_3^2,
\label{HLMGW}
\end{eqnarray}
where we have subtracted a constant term $W(\hat{S}_1^2+\hat{S}_2^2+\hat{S}_3^2)=W\frac{N}{2}\left(\frac{N}{2}+1 \right)$ that does not influence the dynamics. Compared with Eq. (\ref{eqH}), we can see that the model corresponds to a symmetric Foucault pendulum with $W= -(3/16)\omega$ and $\epsilon = -2\Omega$. For large $N$, the Hamiltonian (\ref{HLMGW}) is known to transit through a critical point at $S_0=|\epsilon/W|$ (following, for example, the considerations of \cite{Castanos,Ribeiro2007}): For lower values the linear term dominates and the energy spectrum is smooth. For higher values of $S_0$ the nonlinearity dominates and a discontinuity in the energy spectrum occurs: the change is called a {\em quantum phase transition}. The situation corresponds to the occurrence of a circle of unstable stationary points on the Poincar\'{e} sphere  for $S_0>16 \Omega /(3\omega)$, which follows from Eq. (\ref{S0crit}) if we set $\Delta \omega=0$.

Another special case is  $V=-W$. Relabeling the variables $\hat{S}_1\to \hat{S}_2 \to \hat{S}_3 \to \hat{S}_1$ then leads to the Hamiltonian
\begin{eqnarray}
\hat{H}=\epsilon \hat{S}_1 + 2W \hat{S}_3^2
\end{eqnarray}
that corresponds to an asymmetric pendulum with $\Omega=0,$ $\epsilon =\Delta \omega$ and $2W= (3/16)\omega$. Also here a quantum phase transition occurs, now at $S_0 = |\epsilon/(4 W)|$. For lower $S_0$ the linear term is dominant and the spectrum is smooth; for higher values of $S_0$ a discontinuity in the spectrum occurs corresponding to the occurrence of the unstable stationary point on the Poincar\'{e} sphere for $S_0>8 \Delta \omega /(3\omega)$. Experimental observation of such a bifurcation in collective atomic spins has been reported in \cite{Zibold2010}. We illustrate the correspondence of the quantum and classical models in Fig. \ref{f-Spectra1}.

The general case of both $\Delta \omega \neq 0$ and $\Omega \neq 0$ is not included in the original LMG Hamiltonian (\ref{HLMG}), but can be treated as a generalized LMG model (see, e.g., \cite{Vidal2006,LMG-TO}). In Fig. \ref{f-Spectra2} we show the transition through the critical point with keeping $\Omega$ constant and varying $\Delta \omega$. The effect can be observed in interacting spin systems by applying suitable resonant and off-resonant electromagnetic fields, or in a Foucault pendulum on rotating Earth by varying the asymmetry in the elasticity of the suspension.

Note that when comparing the quantum and classical critical phenomena, $S_0$ was always considered small in the classical model, whereas in the quantum systems the corresponding parameter $N/2$ goes to infinity in the thermodynamic limit. This is the consequence of our choice of the length scale. It is convenient to compare the amplitude of the oscillations with the pendulum length $L$; therefore in the definition of $S_k$ in Eqs. (\ref{Stokes30})---(\ref{Stokes3}) there is $L^2$ in the denominator and $S_k <1$.
If, instead, the length unit were chosen as the size of the vacuum fluctuations $(\hbar/2m\omega)^{1/2}$, then the magnitudes of the dimensionless parameters $S_0$ and $N/2$ would match.

\section{Conclusion}
\label{Sec-Conclusion}
%%%%%%%%%%%%%%%%%%%%%%%%%%%%%%%%%%%%%%%%%%%%%%%%

We have discussed a straightforward analogy between the motion of Foucault-type pendulums and the dynamics of various optical and quantum mechanical physical systems. 
Our approach is based on reducing the dimensionality of the spherical pendulum problem.
Out of the complete 4D phase space of the system we give up one variable: by averaging out the phase, the remaining dynamical variables can be chosen as the three independent Stokes coordinates $S_{1,2,3}$.
There are two integrals of motion which can be visualized  as a sphere and a parabolic cylinder, their lines of  intersection determining the trajectories. The geometric interpretation allows us to identify the boundary between regimes of dominant nonlinearity with unstable stationary points, and linearity-dominated regimes with just two stable stationary points.
 As we checked numerically, the evolution of the reduced model very truly reproduces behavior of the full pendulum model in the regime of $\Omega, \Delta \omega \ll \omega$ and reasonably small amplitudes, $S_0 \lesssim 0.3$.
Deviations from the exact evolution of the pendulum stem from the difference between the actual shape of the orbit and its approximation as an ellipse.

 The classical evolution equations correspond to the quantum mechanical Heisenberg equations of the LMG model with relevance to BEC dynamics or spin squeezing. Even though many interesting quantum phenomena have their counterparts in the pendulum dynamics, there are essential differences as well. Whereas the classical set of equations (\ref{StokesComb}) is closed, the operator equations generated by the quantum Hamiltonian (\ref{eqHQuant}) do not lead to a closed set of equations for their moments: one gets an infinite chain of equations, each coupling to higher order moments. As a result, a quantum system can develop features that are beyond the classical description. Interference structures of Schr\"{o}dinger cat-like states is one of possible examples.  Having in mind the essential differences, our results can be useful in transferring intuition between rather different areas of physics.

%%%%%%%%%%%%%%%%%%%%%%%%%%%%%%%%%%%%%%%%%%%%%%%%%%%%%%
\appendix

\section{Proof of Eq. (\ref{S0crit})}
\label{Sec-Ap1}
%%%%%%%%%%%%%%%%%%%%%%%%%%%%%%%%%%%%%%%%%%%%%%%%
From Eq. (\ref{eqH}) express $S_1$ as a function of $S_3$ and find the radius of the curvature of the parabola as 
$R=(1+S_1^{\prime 2})^{3/2}/|S_1^{\prime \prime}|$, where $S_1^{\prime} = dS_1/dS_3$. We get 
\begin{eqnarray}
R= \frac{8|\Delta \omega|}{3\omega}\left[ 1+ \left( 
\frac{2\Omega}{\Delta \omega} - \frac{3\omega}{8\Delta \omega}S_3
\right)^2 \right]^{3/2} .
\label{ApR}
\end{eqnarray}
For a point on a circle with radius $R$ centered at the line $S_3=0$ we find the relation between the coordinate $S_3$ and the derivative $S_1^{\prime}$ as
\begin{eqnarray}
S_3 = \pm R \frac{S_1^{\prime}}{\sqrt{1+S_1^{\prime 2}}}.
\end{eqnarray} 
Putting equal the radii of the parabola and the circle at a point where the derivatives  $S_1^{\prime}$ of both curves are equal we find
\begin{eqnarray}
S_3 = \frac{8\Delta \omega}{3\omega} \left[ \frac{2\Omega}{\Delta \omega} + \left( 
\frac{2\Omega}{\Delta \omega}\right)^{1/3} \right],
\end{eqnarray} 
which inserted into Eq. (\ref{ApR}) leads to Eq. (\ref{S0crit}) for $S_0=R$.

\section*{Acknowledgment}

%Inspiration from one task of the 2018's International Young Physicists' Tournament  solved by one of us (P. \v{S}.) is acknowledged. 
T.O. is grateful to K. K. Das for stimulating discussions. This work was supported by the   Czech Science Foundation, grant No. 17-20479S.

%%%%%%%%%%%%%%%%%%%%%%%%%%%%%%%%%%%%%%%%%%%%%%%%%%%%%%

\end{document}